\documentclass[lettersize,journal]{IEEEtran} 

\usepackage{color,soul}
\usepackage{subcaption}
\usepackage{graphicx}      
\usepackage{multirow}
\usepackage{comment}
\usepackage{amsmath,amssymb,amsfonts}
\usepackage{amsthm}
\usepackage{tikz}
\usepackage{booktabs}   
\usepackage{array}      
\usepackage{tabularx}   

\usepackage{cite}

\usepackage{mathtools}
\usepackage{bm}
\theoremstyle{definition}

\theoremstyle{remark}
\newtheorem{remark}{Remark}

\title{\LARGE \bf
A Control-Theoretic Foundation for Agentic Systems
}
\author{Ali Eslami$^{1}$ and Jiangbo Yu$^{1}$
	\thanks{$^{1}$Ali Eslami ({\tt\small ali.eslami@mcgill.ca}) and Jiangbo Yu ({\tt\small jiangbo.yu@mcgill.ca}) are with the Department of Civil Engineering, Mcgill University, Montreal, Quebec H3A 0G4, Canada.}
}
\begin{document}
\maketitle
\thispagestyle{plain}
\pagestyle{plain}

\begin{abstract}
This paper develops a control-theoretic framework for analyzing agentic systems embedded within feedback control loops, where an AI agent may adapt controller parameters, select among control strategies, invoke external tools, reconfigure decision architectures, and modify control objectives during operation. These capabilities are formalized by interpreting agency as hierarchical runtime decision authority over elements of the control architecture, leading to an augmented closed-loop representation in which physical states, internal memory, tool outputs, interaction signals, and design variables evolve as a coupled dynamical system. A five-level hierarchy of agency is defined, ranging from fixed control laws to runtime synthesis of control architectures and objectives. The analysis shows that increasing agency introduces interacting dynamical mechanisms such as time-varying adaptation, endogenous switching, decision-induced delays, and structural reconfiguration. The framework is developed in both nonlinear and linear settings, providing explicit design constraints for AI-enabled control systems in safety-critical applications.
\end{abstract}


\section{Introduction}

Artificial intelligence systems are increasingly embedded within feedback
control loops across a wide range of application domains, including
autonomous vehicles, robotics, intelligent transportation systems, and
cyber-physical infrastructure. In these settings, AI components are no
longer confined to perception or planning layers, but actively influence
control decisions during operation. Modern AI-enabled systems can adapt
controller parameters, select among control strategies, invoke external
computational tools, reconfigure decision pipelines, and even modify task
objectives in real time.

This expanded role raises a fundamental question: when an AI system is
granted authority not only to compute control inputs but also to modify
parameters, strategies, tools, architectures, or objectives during
operation, what type of dynamical system must be analyzed?

Classical control theory provides well-established tools for analyzing
systems with fixed controllers~\cite{khalil2002nonlinear}, adaptive
parameters~\cite{ioannou1996robust,aastrom2013computer}, switched
systems~\cite{liberzon2003switching,hespanha1999stability}, and hybrid
dynamical systems~\cite{goebel2009hybrid}. These frameworks typically treat
time variation, switching signals, or structural changes as exogenous or
designer-specified. In contrast, modern AI-driven systems may simultaneously
adapt parameters, switch policies, invoke external computational tools,
reconfigure decision pipelines, and revise task objectives within a single
feedback loop. When these decisions are generated endogenously during
operation, the resulting closed-loop dynamics combine multiple mechanisms
whose interaction must be explicitly characterized.

In this work, we focus on \emph{runtime agency}, namely decision authority
exercised by the AI during closed-loop operation. Design-time choices, such
as admissible controller families, tool libraries, and governance
constraints, are treated as fixed prior to deployment. The objective is to
characterize how runtime decision authority alters the dynamical structure
of the closed loop and what implications this has for stability and safety.

This problem is increasingly relevant across application domains. In
robotics, language-model-based agents reason about tasks, select skills, and
orchestrate collections of controllers~\cite{lima2026agentic}. In
transportation systems, agentic vehicles adapt behavior in complex and
uncertain environments~\cite{yu2025agentic}. Broader analyses highlight that
increasing agency introduces operational and governance challenges~\cite{yu2025preparing},
while recent security studies show that agentic decision-making can create
vulnerabilities across cognitive, software, and physical
layers~\cite{eslami2025security}.

Related work on formal methods for safety-critical systems provides
correct-by-construction guarantees by synthesizing controllers that satisfy
formal specifications \cite{yin2024formal}, including abstraction-based methods \cite{tabuada2009verification}, temporal logic
synthesis \cite{belta2017temporal}, and optimization-based formulations \cite{belta2019formal}.
These approaches typically assume that system specifications, control
architectures, and objective functions remain fixed at design time.

At the same time, recent work on LLM-assisted control has explored a variety
of integration patterns, including interactive tuning and controller
configuration~\cite{tohma2025smartcontrol}, adaptive compensation and parameter
adaptation~\cite{zhou2025llms,zahedifar2025llm}, model predictive
control-based decision support~\cite{wu2025instructmpc}, specification and
controller synthesis from natural language~\cite{li2025natural,bayat2025llm},
and supervisory coordination in complex systems such as manufacturing and
industrial automation \cite{lim2025large,xia2025control,narimani2025agenticcontrol,maher2025llmpc,aydin2026mrac}.
While these works demonstrate promising capabilities, they primarily study
individual mechanisms in isolation.

Recent advances in safe reinforcement learning have integrated Lyapunov and
barrier functions to provide stability and safety certificates during
learning and deployment~\cite{Kushwaha2026}, complementing
classical RL methods that often lack explicit
guarantees~\cite{Sutton2018}. By contrast, we treat any
endogenous parameter, objective, or architectural update as part of the
closed-loop dynamics and derive stability constraints that can be
superimposed on learning-based designs.

The recent proliferation of LLM-based agents has produced systems capable of
memory, tool use, and planning~\cite{Xi2023}, with robotic
applications such as SayCan~\cite{Ahn2022}. While these works illustrate
the mechanisms we model, such as strategy selection, tool invocation, and
reconfiguration, they focus on empirical performance rather than closed-loop
stability. Our framework complements them by providing a unified dynamical
model of agentic decision-making and stability conditions for safety-critical
deployment.

Furthermore, the central premise of this paper is that agency can be interpreted as
runtime decision authority over elements of the control architecture.
We
characterize agency through the variables that an agent is allowed to modify
during operation, including controller parameters, controller families,
workflow configurations, and control objectives. This interpretation leads
to a hierarchy of agency levels, following the conceptual classification
in~\cite{yu2025preparing}, in which increasing authority corresponds to
progressively richer classes of dynamical systems.

From a control-theoretic perspective, this interpretation alters the
structure of the closed-loop dynamics. When multiple forms of runtime
decision authority are exercised simultaneously, the resulting system
exhibits coupled effects involving adaptation, switching, delay, and
structural reconfiguration.

To analyze these effects, we introduce an augmented closed-loop
representation that incorporates both the physical state and the internal
decision variables of the agent. Within this framework, we show that
stability depends critically on the coupling between adaptation,
switching, delay, and structural reconfiguration, leading to explicit
conditions linking adaptation rates, switching frequencies, and delay
bounds.

Therefore, the contribution of this paper is to develop a unified control-theoretic
framework for modeling agentic decision-making as an intrinsic component of
closed-loop dynamics and to characterize the resulting coupling-induced
stability constraints. This perspective provides a principled basis for
analyzing and governing AI-enabled control systems in safety-critical
applications.

The main contributions are summarized as follows:
\begin{itemize}
	\item A unified dynamical formulation of agentic control systems in which
	memory, learning, tool invocation, interaction, and goal descriptors are
	incorporated within a single closed-loop architecture, with agency
	interpreted as runtime decision authority.
	
	\item A control-theoretic interpretation of the agency hierarchy
	of~\cite{yu2025preparing}, relating increasing levels of authority to
	progressively richer dynamical classes, including time-varying,
	switched, hybrid, and online controller design dynamics.
	
	\item An analysis of how coupled mechanisms, including parameter
	adaptation, endogenous switching, decision-induced delays, and structural
	reconfiguration, jointly influence closed-loop stability.
	
	\item Explicit stability conditions that relate adaptation rates,
	switching frequencies, and delay bounds, providing design constraints for
	agentic control systems.
\end{itemize}

The remainder of the paper is organized as follows. Section~II presents the
unified agentic control architecture. Section~III introduces the agency
hierarchy. Section~IV analyzes the coupled dynamics of adaptation,
switching, delay, and reconfiguration while Section~V discusses the implications
for design and governance. Section~VI presents simulation examples and,
Section~VII concludes and outlines future directions.

\section{Unified Agentic Control Architecture}
\subsection{General Nonlinear Formulation}
Consider a physical system
\begin{equation}
	\dot{x}(t)=f\big(x(t),u(t),w(t)\big), \qquad
	y(t)=h\big(x(t),w(t)\big),
	\label{eq:plant}
\end{equation}
where $x(t)$ is the state, $u(t)$ the control input, $y(t)$ the observation, and $w(t)$ exogenous disturbances. 
The objective of this section is to embed agentic AI capabilities within a unified closed-loop representation and to interpret agency as runtime decision authority over elements of the control architecture.

Agentic AI systems exhibit capabilities such as reasoning, memory and context accumulation, retrieval-augmented information access (RAG), tool use through external APIs, interaction with humans or other agents, and dynamic goal specification. These capabilities are represented through a set of variables summarized in Table~\ref{tab:agentic_capabilities}, which define the signals and states available to the controller.

\begin{table}[t]
	\caption{Agentic capabilities and their representation in the control architecture.}
	\label{tab:agentic_capabilities}
	\centering
	\renewcommand{\arraystretch}{1.05}
	\begin{tabularx}{\linewidth}{l l X}
		\toprule
		\textbf{Capability} & \textbf{Variable} & \textbf{Role in Architecture} \\
		\midrule
		
		Memory / context 
		& $m(t)$ 
		& Internal state summarizing past information and supporting reasoning \\
		
		Adaptation / policy update 
		& $\theta(t)$ 
		& Evolution of internal policy or controller parameters \\
		
		Tool use / retrieval 
		& $z(t)$ 
		& Outputs of external tools or APIs (e.g., planners, predictors, retrieval modules) \\
		
		Interaction 
		& $r(t)$ 
		& Inputs from humans or other agents, possibly with ambiguity \\
		
		Goal specification 
		& $\zeta(t)$ 
		& Descriptor shaping the control objective and task priorities\\
		\bottomrule
	\end{tabularx}
\end{table}

Define the information state
\begin{equation}
	I(t)=\big\{y(t),\,m(t),\,z(t),\,r(t)\big\},
	\label{eq:info_set}
\end{equation}
which aggregates observations ($y(t)$), internal memory ($m(t)$), retrieved information ($z(t)$), and interaction signals ($r(t)$).

The memory state evolves as
\begin{equation}
	\dot{m}(t)=\Phi_m\big(m(t),y(t),z(t),r(t)\big),
	\label{eq:memory_dyn}
\end{equation}
capturing reasoning, belief updates, and context accumulation.

External tool outputs $z(t)$ include predictions, plans, or retrieved information. These processes often incur computation or communication delays, modeled as
\begin{equation}
	z(t)=\mathcal{T}\big(I(t-\tau_z(t))\big),
	\label{eq:tool_delay}
\end{equation}
where $\tau_z(t)\ge 0$ denotes tool or retrieval latency.

Interaction signals $r(t)$ represent inputs from humans or other agents. When language-mediated,
\[
r(t)=\bar r(t)+\Delta(t),
\]
where $\Delta(t)$ captures semantic ambiguity which is set to zero in this paper and is left for future studies.

The goal descriptor $\zeta(t)$ parameterizes the control objective:
\begin{equation}
	\zeta(t)=\Gamma\big(I(t-\tau_\zeta(t))\big),
	\label{eq:goal_gen}
\end{equation}
allowing for delays in goal inference or re-planning.

Controller parameters evolve according to
\begin{equation}
	\dot{\theta}(t)=\Phi_\theta\big(\theta(t),I(t-\tau_\theta(t))\big),
	\label{eq:theta_dyn}
\end{equation}
capturing learning or adaptation with possible update latency.

The control input is generated as
\begin{equation}
	u(t)=\pi_{\alpha(t)}\big(I(t-\tau_u(t));\theta(t),\zeta(t)\big),
	\label{eq:controller_general}
\end{equation}
where $\alpha(t)$ denotes the controller index or policy class, selecting a control law from a predefined family, and $\tau_u(t)$ represents computation or inference delay in the decision process.

Define the design state
\begin{equation}
	d(t)=\big(\zeta(t),\,\alpha(t),\,c(t)\big),
	\label{eq:design_state}
\end{equation}
which collects the variables shaping the controller structure and objective, where $c(t)$ denotes the control architecture or workflow configuration, specifying how components such as estimation, planning, and control are interconnected.

Agency is determined by which components of
\begin{equation}
	\{\theta(t),\,\alpha(t),\,c(t),\,\zeta(t)\}
	\label{eq:agency_vars}
\end{equation}
are fixed, selectable, or generated during operation.

Note that the goal descriptor $\zeta(t)$ can also be interpreted as encoding reference trajectories or task specifications generated online. In this sense, agentic systems extend classical reference tracking and trajectory planning frameworks by allowing references to evolve dynamically based on internal reasoning, external tools, or interaction signals. This introduces an additional layer of dynamics that couples planning and control within the closed loop.

\subsection{Linear Specialization}
\label{sec:linear}

For a linear plant
\begin{equation}
	\dot{x}(t)=Ax(t)+Bu(t)+Ew(t), \qquad
	y(t)=Cx(t)+Dw(t),
\end{equation}
the memory dynamics become
\begin{equation}
	\dot{m}(t)=A_m m(t)+B_m y(t)+G_m z(t)+H_m r(t),
\end{equation}
with $A_m$ Hurwitz. Define the linear information set as
\begin{equation}
	\tilde{\chi}(t)=
	\begin{bmatrix}
		y(t)\\
		m(t)\\
		z(t)\\
		r(t)
	\end{bmatrix}.
\end{equation}
The cost function is
\begin{equation}
	J=\int_{0}^{\infty}\ell_{\zeta(t)}\big(x(t),u(t)\big)\,dt,
\end{equation}
and the controller is
\begin{equation}
	u(t)=K_{\alpha(t)}\big(\theta(t),\zeta(t)\big)\tilde{\chi}(t-\tau_u(t)),
\end{equation}
recovering fixed, adaptive, switched, or modular control depending on the agency level.

\subsection{Definition of Agency}

Agency is defined as runtime decision authority over elements of the control architecture. Let
\[
q(t)=\{\theta(t),\alpha(t),c(t),\zeta(t)\}
\]
denote the decision variables governing parameters, controller selection, workflow composition, and objectives. Agency corresponds to the subset of $q(t)$ whose evolution is determined endogenously.
Note that signals such as $m(t)$, $z(t)$, and $r(t)$ influence the decision process but are not themselves decision variables.

Furthermore, it should be noted that autonomy and agency are related but distinct concepts. 
Autonomy refers to the ability of a system to generate control inputs without external intervention, given a fixed control architecture. In this case, the mapping from information to action is predetermined, and the system operates independently within that structure.
Agency, in contrast, refers to the ability of the system to modify the elements that define this mapping during operation. Specifically, an agentic system can influence the evolution of the decision variables $q(t)=\{\theta(t),\alpha(t),c(t),\zeta(t)\}$, thereby altering how control inputs are generated.
Thus, autonomy operates at the level of execution within a fixed policy, whereas agency operates at the level of policy formation and modification. A system may be fully autonomous yet non-agentic if its control law is fixed, while a system exhibits agency when it has runtime authority to adapt parameters, switch strategies, reconfigure workflows, or revise objectives.

\section{Agency Levels as Hierarchical Runtime Authority}
\label{sec:hierarchy}

The framework of Section II defines a set of decision variables $q(t)=\{\theta(t),\alpha(t),c(t),\zeta(t)\}$ whose evolution shapes the closed-loop system. The agency hierarchy characterizes how granting runtime authority over these variables alters the resulting dynamics.
The hierarchy is cumulative: higher levels inherit the capabilities of lower levels while introducing additional endogenous variables, switching laws, structural changes, or design dynamics into the closed loop, together with their associated delays.
From an AI perspective, this corresponds to increasing runtime discretion. From a control perspective, it corresponds to increasing dynamical complexity: fixed systems, time-varying systems, switched systems, hybrid systems, and hybrid systems with endogenous design evolution.

\subsection{Hierarchy Description}

\textbf{Level 1} (Fixed policy execution):  
All decision variables are fixed:
\begin{equation}
	\theta(t)\equiv \bar\theta,\qquad
	d(t)\equiv \bar d.
\end{equation}
The controller is
\begin{equation}
	u(t)=\pi_{\bar\alpha}\big(I(t-\tau_u(t));\bar\theta,\bar\zeta\big),
\end{equation}
or follows a predefined switching rule. The resulting system is fixed (or exogenously switched), with delay arising only from sensing or computation.

\textbf{Level 2} (Internal adaptation and memory):  
Internal states become endogenous:
\begin{equation}
	u(t)=\pi_{\bar\alpha}\big(I(t-\tau_u(t));\theta(t),\bar\zeta\big),
\end{equation}
\begin{align*}
	\dot{\theta}(t)&=\Phi_\theta\big(\theta(t) ,I(t-\tau_\theta(t))\big), \\
	\dot m(t)&=\Phi_m\big(m(t),y(t),z(t),r(t)\big)
\end{align*}
The addition of $(\theta,m)$ yields a time-varying system. This level captures reasoning, memory, and context accumulation, with delays arising from inference and adaptation.

\textbf{Level 3} (Tool selection and strategy switching):  
Runtime authority expands to selecting among predefined alternatives:
\begin{equation}
	u(t)=\pi_{\alpha(t)}\big(I(t-\tau_u(t));\theta_{\alpha(t)}(t),\zeta_{g(t)}\big),
\end{equation}
\begin{equation}
	(\alpha(t),g(t))=\Sigma\big(I(t-\tau_\sigma(t))\big).
\end{equation}
The system becomes switched with endogenous decision delay $\tau_\sigma(t)$. This level corresponds to tool use and strategy selection, where switching frequency and latency jointly influence stability. For example, an agent may switch between a nominal feedback controller and a planning-based controller by invoking a trajectory optimization tool when predicted tracking error exceeds a threshold.

\textbf{Level 4} (Dynamic workflow and tool orchestration):  
The agent can modify the interconnection structure:
\begin{equation}
	u(t)=\Pi\big(I(t-\tau_u(t));\theta(t),\zeta(t),c(t)\big),
\end{equation}
\begin{equation}
	z(t)=\mathcal T_{c(t)}\big(I(t-\tau_z(t))\big).
\end{equation}
The variable $c(t)$ determines workflow composition. For example, the agent may switch between a direct state-feedback pipeline and a perception–estimation–planning–control pipeline by inserting or removing modules (e.g., adding a state estimator and a trajectory planner), thereby changing the signal flow and internal dynamics.

\textbf{Level 5} (Goal re-planning and generative design):  
The design tuple becomes endogenous:
\begin{equation}
	d(t)=(\zeta(t),\alpha(t),c(t))\in\mathcal G_c,
\end{equation}
\begin{equation}
	\dot d(t)=\Omega\big(I(t-\tau_d(t)),d(t)\big).
\end{equation}
The system evolves on multiple timescales with supervisory delay $\tau_d(t)$, capturing goal re-planning and generative decision-making.

To connect these levels with control-theoretic concepts, Table~\ref{tab:agentic_control_mapping} maps common agentic AI mechanisms to their dynamical interpretations.

\begin{table*}[t]
	\caption{Control-theoretic interpretation of common agentic AI mechanisms.}
	\label{tab:agentic_control_mapping}
	\centering
	\renewcommand{\arraystretch}{1}
	\begin{tabularx}{\linewidth}{l c X X}
		\toprule
		\textbf{AI Mechanism} & \textbf{Level} & \textbf{Control Interpretation} & \textbf{Engineering Implication} \\
		\midrule
		
		Tool use / API calls 
		& L3 
		& Hybrid interaction (switching + exogenous input + dynamic extension) 
		& Introduces discrete events, auxiliary states, and tool-dependent latency $\tau_z$; requires hybrid and robustness analysis \\
		
		Reasoning / internal inference 
		& L2 
		& Internal state evolution with computation delay 
		& Adds latent states and inference delay $\tau_\theta$; may reduce stability margins \\
		
		Retrieval (RAG) 
		& L2 
		& Exogenous information injection / measurement augmentation 
		& Alters available information without changing controller structure; may introduce retrieval latency $\tau_z$ and delayed inputs \\
		
		Memory / context accumulation 
		& L2 
		& Internal state augmentation 
		& Introduces persistent internal states influencing decisions; may cause drift if unbounded \\
		
		Dynamic tool / workflow orchestration 
		& L4 
		& Structural reconfiguration / hybrid system 
		& Changes interconnection structure and tool pathways; stability must hold under reconfiguration and delay \\
		
		Goal re-planning 
		& L5 
		& Objective-function evolution / cost reshaping 
		& Alters optimization landscape and Lyapunov structure; requires admissible goal constraints \\
		
		\bottomrule
	\end{tabularx}
\end{table*}
Furthermore, Interaction signals $r(t)$ are treated as exogenous inputs influencing the decision process and are therefore not included as a separate row.

These mechanisms introduce switching, delay, state augmentation, and objective variation into the closed loop. Their interaction, particularly the coupling of switching and latency, plays a central role in determining stability.

\section{Stability Implications}
\label{sec:stability}

Sections~II and~III introduced a unified representation of agentic control
systems and a hierarchy of runtime decision authority. We now analyze how
these decision mechanisms affect the closed-loop dynamics. The main point is
that agentic systems do not merely introduce isolated effects such as
adaptation, switching, delay, or structural reconfiguration. Rather, these
mechanisms coexist and interact. Stability must therefore be understood in
terms of the joint evolution of the endogenous decision variables and
their associated delays.

\subsection{Augmented Closed-Loop Representation}
Let
\begin{equation}
	\xi(t):=\bigl(x(t),\theta(t),\sigma(t),c(t),\zeta(t),m(t)\bigr)
	\label{eq:augmented_state}
\end{equation}
denote the augmented state, where \(x\) is the plant state, \(m\) the internal
memory state, \(\theta\) the adaptable parameter state, \(\sigma\) the active
controller or mode index, \(c\) the workflow configuration, and \(\zeta\) the
goal descriptor. Define the associated design tuple
\begin{equation}
	d(t):=\bigl(\zeta(t),\sigma(t),c(t)\bigr).
\end{equation}

Using the notation of Section~II, the continuously evolving components satisfy
\begin{align}
	\dot{x}(t) &= f\bigl(x(t),u(t),w(t)\bigr), \label{eq:aug_x}\\
	\dot{m}(t) &= \Phi_m\bigl(m(t),y(t),z(t),r(t)\bigr), \label{eq:aug_m}\\
	\dot{\theta}(t) &= \Phi_\theta\bigl(\theta(t),I(t-\tau_\theta(t))\bigr),
	\label{eq:aug_theta}
\end{align}
while the goal descriptor is generated according to
\begin{equation}
	\zeta(t)=\Gamma\bigl(I(t-\tau_\zeta(t))\bigr).
	\label{eq:aug_zeta}
\end{equation}
The control law is
\begin{equation}
	u(t)=\pi_{\sigma(t)}\bigl(I(t-\tau_u(t));\theta(t),\zeta(t)\bigr),
	\label{eq:aug_control}
\end{equation}
and discrete updates are introduced through
\begin{equation}
	\sigma^+=\Omega_\sigma\bigl(I(t-\tau_\sigma(t)),q(t)\bigr),
	\qquad
	c^+=\Omega_c\bigl(I(t-\tau_c(t)),q(t)\bigr).
	\label{eq:aug_jumps}
\end{equation}

This representation shows that increasing agency introduces four interacting
mechanisms into the closed loop: parameter evolution, endogenous switching,
structural reconfiguration, and delay. Each mechanism is well studies in isolation in classical control literature, but their simultaneous presence and endogenous nature leads to coupled stability
constraints that do not appear when they are analyzed separately.

\subsection{Baseline Stability of Isolated Agency Mechanisms}

When considered individually, the main agency mechanisms admit standard
stability certificates. Slowly varying parameter adaptation can be treated
using Lyapunov arguments for time-varying systems when the adaptation rate is
sufficiently small~\cite{ioannou1996robust}. Mode switching can be
certified using multiple Lyapunov functions together with average dwell-time
conditions~\cite{hespanha2004uniform,liberzon2003switching}. Architectural reconfiguration can
be analyzed as a hybrid jump process with bounded jump growth and sufficiently
slow reconfiguration~\cite{goebel2009hybrid}. The central difficulty in agentic
systems is therefore not the presence of any one mechanism in isolation, but
their coexistence and mutual interaction.

\subsection{Cross-Coupling Between Adaptation and Switching}

The most immediate interaction arises between Level~2 adaptation and Level~3
switching. Either mechanism may be stabilizing in isolation, yet their
coupling can destabilize the closed loop if switching reacts too quickly to an
adapting parameter state.

Consider the coupled Level~2-Level~3 dynamics
\begin{align}
	\dot{x}(t) &= f_{\sigma(t)}\bigl(x(t),\theta(t)\bigr), \label{eq:l23_x}\\
	\dot{\theta}(t) &= \Phi_\theta\bigl(\theta(t),I(t-\tau_\theta(t))\bigr),
	\label{eq:l23_theta}\\
	\sigma^+ &= \Omega_\sigma\bigl(I(t-\tau_\sigma(t)),\theta(t)\bigr),
	\qquad \sigma(t)\in\mathcal{P}.
	\label{eq:l23_sigma}
\end{align}

\textbf{Assumption 1}:
	\label{ass:Vp}
	The admissible parameter set \(\Theta\) is compact and forward invariant
	under \(\Phi_\theta\). For each mode \(p\in\mathcal{P}\), there exists a
	continuously differentiable Lyapunov function \(V_p(x,\theta)\) and
	constants \(\alpha_1,\alpha_2,\gamma>0\) such that
	\begin{equation}
		\alpha_1\|x\|^2\le V_p(x,\theta)\le \alpha_2\|x\|^2
		\label{eq:Vp_bounds}
	\end{equation}
	for all admissible \(\theta\in\Theta\), and
	\begin{equation}
		\frac{\partial V_p}{\partial x}(x,\theta)\,f_p(x,\theta)
		\le -\gamma V_p(x,\theta)
		\label{eq:Vp_frozen_decay}
	\end{equation}
	for frozen \(\theta\).

\textbf{Assumption 2}:
	There exists \(L_\theta>0\) such that
	\begin{equation}
		\|
		\frac{\partial V_p}{\partial\theta}(x,\theta)\,
		\Phi_\theta(\theta,I) \|	
		\le L_\theta\rho\,V_p(x,\theta)
		\label{eq:cross_perturb_bound}
	\end{equation}
	for all \(p\in\mathcal{P}\), all admissible \(\theta\in\Theta\), and all
	\(I\), where
	\begin{equation}
		\|\Phi_\theta(\theta,I)\|\le\rho
		\label{eq:theta_rate}
	\end{equation}
	for some \(\rho>0\).

\textbf{Assumption 3}:
	There exists \(\nu\ge 1\) such that
	\begin{equation}
		V_p(x,\theta)\le\nu\,V_{p'}(x,\theta),
		\qquad
		\forall x,\ \forall\theta\in\Theta,\ \forall p,p'\in\mathcal{P}.
		\label{eq:Vp_compare}
	\end{equation}

The following theorem analyzes the coupling between adaptation and switching in levels 2 and 3 and its proof is provided in Appendix \ref{Appendix:1}.

\textbf{Theorem 1}:
	Suppose Assumptions~1-3 hold and
	\begin{equation}
		\gamma > L_\theta\rho.
		\label{eq:positive_margin}
	\end{equation}
	If the switching signal satisfies the average dwell-time condition
	\begin{equation}
		N_\sigma(t,t+T)\le N_0+\frac{T}{\tau_a},
	\end{equation}
	with
	\begin{equation}
		\tau_a > \frac{\ln\nu}{\gamma-L_\theta\rho},
		\label{eq:cross_condition}
	\end{equation}
	then the plant state \(x(t)\) converges exponentially to the origin and
	\(\theta(t)\) remains bounded.

\begin{remark}
	\label{rem:cross_coupling_interp}
	Note that the adaptation rate \(\rho\) reduces the
	effective decay margin from \(\gamma\) to \(\gamma-L_\theta\rho\), thereby
	tightening the allowable switching condition. Faster adaptation therefore
	requires slower switching.
\end{remark}

\begin{remark}
	\label{rem:rationality}
	Note that a decision rule may be locally rational in the sense of always selecting the
	mode that appears most effective under the current information state.
	However, if this produces excessive switching while the internal parameter
	state is still adapting, the closed loop may become unstable. Stability
	therefore requires temporal regularity constraints (e.g., dwell-time, hysteresis, or bounded update rates).
\end{remark}

\subsection{Delay-Switching Coupling}

In agentic systems, delay is rarely introduced by a single source. Instead,
multiple latencies accumulate across sensing, reasoning, tool invocation,
switching, and supervisory reconfiguration. Define the total effective delay
\begin{equation}
	\tau_{\mathrm{tot}}(t)
	:=
	\tau_u(t)+\tau_\theta(t)+\tau_z(t)+\tau_\sigma(t)+\tau_c(t)+\tau_\zeta(t),
	\label{eq:tau_total}
\end{equation}
and let
\begin{equation}
	\bar{\tau}:=\sup_{t\ge 0}\tau_{\mathrm{tot}}(t).
\end{equation}

These delays interact with endogenous switching decisions, altering the
effective stability margin of the closed-loop system. The resulting behavior
cannot be characterized by analyzing delay or switching independently.

\textbf{Assumption 4}:
	For each mode \(p\in\mathcal{P}\), there exists a Lyapunov-Krasovskii
	functional \(W_p\) defined on \(C([-\bar{\tau},0],\mathbb{R}^n)\) and
	constants \(\underline{\alpha},\overline{\alpha},\gamma,\beta>0\) such that
	\begin{equation}
		\underline{\alpha}\|\phi(0)\|^2
		\le W_p(\phi)
		\le\overline{\alpha}\|\phi\|^2_{[-\bar{\tau},0]},
	\end{equation}
	and along flows in mode \(p\),
	\begin{equation}
		\dot{W}_p(x_t)\le-(\gamma-\beta\bar{\tau})\,W_p(x_t).
		\label{eq:LK_decay}
	\end{equation}
	There also exists \(\nu\ge 1\) such that
	\[
	W_p(\phi)\le\nu\,W_{p'}(\phi),
	\qquad \forall \phi,\ \forall p,p'\in\mathcal{P}.
	\]

\textbf{Proposition 1}:
	Suppose Assumption~4 holds and
	\begin{equation}
		\gamma>\beta\bar{\tau}.
	\end{equation}
	If the switching signal satisfies the average dwell-time condition
	\[
	N_\sigma(t,t+T)\le N_0+\frac{T}{\tau_a},
	\]
	with
	\begin{equation}
		\tau_a > \frac{\ln\nu}{\gamma-\beta\bar{\tau}},
		\label{eq:delay_adt}
	\end{equation}
	then the origin of the delayed switched system is exponentially stable.

\begin{proof}
	Assumption~4 gives exponential decay of \(W_p\) along flows with
	effective rate \(\gamma-\beta\bar{\tau}\). At switching instants, the
	comparability condition yields a multiplicative jump bound of \(\nu\).
	Combining these two effects with the average dwell-time condition gives
	\[
	W(x_{t+T})
	\le
	\nu^{N_0}
	\exp\!\left(
	-\Bigl(\gamma-\beta\bar{\tau}-\tfrac{\ln\nu}{\tau_a}\Bigr)T
	\right)
	W(x_t),
	\]
	which decays exponentially under~\eqref{eq:delay_adt}. \hfill
\end{proof}

\begin{remark}
	\label{rem:delay_margin}
	Note that when both switching and delay are present, delay reduces the
	effective decay rate from \(\gamma\) to \(\gamma-\beta\bar{\tau}\), thereby
	tightening the admissible dwell-time. From an agentic perspective,
	cumulative latencies due to reasoning, retrieval, and tool invocation are
	therefore stability-critical quantities: increasing delay necessitates more
	conservative switching policies.
\end{remark}

\subsection{Fully Coupled Multi-Mechanism Dynamics}
\label{sec:fully_coupled}
To avoid notational ambiguity, we distinguish between the full augmented
state introduced in~\eqref{eq:augmented_state} and the reduced state used
in the Lyapunov analysis. In this subsection, we define
\[
\chi := (\bar{\xi},\theta,d),
\qquad
\bar{\xi} := (x,m),
\]
where $\bar{\xi}$ denotes the continuously evolving plant-memory substate.
The continuously evolving goal descriptor $\zeta$ is absorbed into $d$,
which collects all continuously evolving design variables. The discrete
variables $(\sigma,c)$ enter only through the configuration index
$p(t)=(\sigma(t),c(t))$. The following theorem provides the stability analysis for the fully coupled system and its proof is provided in Appendix \ref{Appendix:2}.

\textbf{Theorem 2}:
	Consider the augmented hybrid closed-loop system with state
	\[
	\chi(t):=(\bar{\xi}(t),\theta(t),d(t)),
	\qquad
	\bar{\xi}(t):=(x(t),m(t)),
	\]
	where \(x\) is the plant state, \(m\) is the agent memory
	state, \(\theta\) denotes continuously adapted controller parameters, and
	\(d\) collects continuously evolving endogenous design variables. Let the
	discrete configuration be
	\[
	p(t):=(\sigma(t),c(t))\in\mathcal{P}\times\mathcal{C},
	\]
	and let \(\{t_k\}_{k\ge 1}\) denote the ordered jump times. Assume the system
	is complete (solutions exist for all $t\ge 0$) and Zeno-free.
	
	For each \(p\in\mathcal{P}\times\mathcal{C}\), suppose there exists a
	continuously differentiable Lyapunov function
	\(V_p:\mathbb{R}^N\to\mathbb{R}_{\ge 0}\) (with \(N=\dim(\chi)\)) and
	constants \(\underline{\alpha},\overline{\alpha}>0\) such that
	\[
	\underline{\alpha}\|\chi\|^2
	\le V_p(\chi)
	\le \overline{\alpha}\|\chi\|^2,
	\qquad \forall \chi,\ \forall p.
	\tag{A1}
	\]
	
	Assume further that the following conditions hold.
	
	\textit{(A2) Frozen decay.}
	Along any flow interval on which \(p(t)\equiv p\) is constant, the
	plant-memory dynamics with frozen \(\theta\) and \(d\),
	\[
	\dot{\bar{\xi}}=F_p(\bar{\xi};\theta,d),
	\]
	satisfy
	\[
	\frac{\partial V_p}{\partial \bar{\xi}}F_p(\bar{\xi};\theta,d)
	\le -\gamma V_p(\chi)
	\tag{A2}
	\]
	for some \(\gamma>0\).
	
	\textit{(A3) Adaptation sensitivity.}
	There exist constants \(L_\theta,\rho\ge 0\) such that
	\[
	\left|
	\frac{\partial V_p}{\partial \theta}\,
	\Omega_\theta(\theta,I)
	\right|
	\le L_\theta \rho\, V_p(\chi),
	\qquad
	\|\Omega_\theta(\theta,I)\| \le \rho.
	\tag{A3}
	\]
	
	\textit{(A4) Design-evolution sensitivity.}
	There exist constants \(L_d,\eta\ge 0\) such that
	\[
	\left|
	\frac{\partial V_p}{\partial d}\,
	\Omega_d(d,I)
	\right|
	\le L_d \eta\, V_p(\chi),
	\qquad
	\|\Omega_d(d,I)\| \le \eta.
	\tag{A4}
	\]
	
	\textit{(A5) Delay penalty.}
	The total effective delay satisfies
	\[
	\tau_{\mathrm{tot}}(t)\le \bar{\tau},
	\]
	and the delay contribution to the Lyapunov derivative is bounded by
	\[
	\Delta_\tau V_p(\chi)\le \beta \bar{\tau}\, V_p(\chi)
	\tag{A5}
	\]
	for some \(\beta \ge 0\).
	
	\textit{(A6) Jump comparability.}
	At each jump time \(t_k\),
	\[
	V_{p(t_k^+)}(\chi(t_k^+))
	\le
	\nu_\sigma^{\delta_{\sigma,k}}
	\nu_c^{\delta_{c,k}}
	V_{p(t_k^-)}(\chi(t_k^-)),
	\tag{A6}
	\]
	where \(\delta_{\sigma,k},\delta_{c,k}\in\{0,1\}\) indicate controller
	switching and architectural reconfiguration, respectively, and
	\(\nu_\sigma,\nu_c\ge 1\).
	
	\textit{(A7) Average dwell-time bounds.}
	For all \(t\ge 0\) and \(T\ge 0\),
	\[
	N_\sigma(t,t+T)\le N_{0,\sigma}+\frac{T}{\tau_{a,\sigma}},
	\qquad
	N_c(t,t+T)\le N_{0,c}+\frac{T}{\tau_{a,c}}.
	\tag{A7}
	\]
	
	Define the effective decay margin
	\[
	\lambda := \gamma - L_\theta \rho - L_d \eta - \beta \bar{\tau}.
	\tag{A8}
	\]
	
	If
	\[
	\lambda > 0
	\quad \text{and} \quad
	\lambda >
	\frac{\ln \nu_\sigma}{\tau_{a,\sigma}}
	+
	\frac{\ln \nu_c}{\tau_{a,c}}
	\tag{A9}
	\label{eq:fully_coupled_condition}
	\]
	then there exist constants \(K>0\) and \(\omega>0\) such that
	\[
	\|\chi(t)\|\le K e^{-\omega t}\|\chi(0)\|,
	\qquad \forall t\ge 0.
	\]
	Therefore, the reduced state \(\chi(t)\) converges exponentially to the
	origin, and hence the plant state \(x(t)\), as a component of \(\chi(t)\),
	also converges exponentially to the origin.

\begin{remark}
	Theorem~2 provides a conservative sufficient condition
	rather than a complete characterization of arbitrary fully coupled agentic
	dynamics. Its role is to show that adaptation, delay, controller switching,
	architectural reconfiguration, and endogenous design evolution can be
	certified simultaneously under a common Lyapunov framework. Furthermore, Table \ref{Table:III} demonstrates the stability constraint and the design rules that are necessary to consider when increasing the levels of agency.
\end{remark}

\begin{remark}
	\label{rem:mas}
	The stability condition \eqref{eq:fully_coupled_condition} depends jointly on quantities associated
	with different decision layers, including adaptation rates
	\((L_\theta,\rho)\), delay bounds \(\bar{\tau}\), and switching and
	reconfiguration parameters \((\nu_\sigma,\nu_c,\tau_{a,\sigma},\tau_{a,c})\).
	In a multi-agent architecture, these quantities may be governed by distinct
	agents (e.g., learning modules, supervisory controllers, or planning units)
	with no shared internal state. As a result, the stability condition cannot, in
	general, be verified or enforced by any single agent in isolation.
	This reveals an inherent coordination requirement: stability must be
	maintained through either a shared stability budget across agents or a
	supervisory mechanism that regulates their joint behavior. Another approach is to maintain stability through coordination on the stability budget and this idea is left for our future works.
\end{remark}

\begin{table}[t]
	
	\caption{Stability constraints induced by agentic mechanisms across agency levels. Higher levels introduce additional constraints and coupling effects.}
	\centering
	\label{Table:III}
	\small
	\begin{tabular}{lll}
		\toprule
		Level & Constraint & Design rule \\
		\midrule
		
		L2 
		& $L_\theta \rho < \gamma$ 
		& limit adaptation rate \\
		\midrule
		L3 
		& $\tau_a > \dfrac{\ln \nu}{\gamma - L_\theta \rho}$ 
		& enforce dwell-time \\
		\midrule
		\begin{tabular}[c]{@{}l@{}}
			L3 + \\
			delay
		\end{tabular}
		& $\begin{aligned}
		\tau_a &> \dfrac{\ln \nu}{\gamma - \beta \bar{\tau}}
		\end{aligned}$
		& \begin{tabular}[c]{@{}l@{}}
			reduce switching \\
			under latency
		\end{tabular} \\
		\midrule
		L4 
		& $\tau_{a,c} > \dfrac{\ln \nu_c}{\gamma}$ 
		& \begin{tabular}[c]{@{}l@{}}
			limit reconfiguration \\
			rate
		\end{tabular} \\
		\midrule
		L5
		& $\begin{aligned}
		\gamma >\;& L_\theta \rho + L_d \eta + \beta \bar{\tau} \\
		&+ \dfrac{\ln \nu_\sigma}{\tau_{a,\sigma}} 
		+ \dfrac{\ln \nu_c}{\tau_{a,c}}
		\end{aligned}$
		& \begin{tabular}[c]{@{}l@{}}
			allocate shared \\
			stability budget
		\end{tabular} \\
		
		\bottomrule
	\end{tabular}
\end{table}
\section{Discussion}
\label{sec:discussion}

\subsection{Interpretation and Relation to Existing Work}
\label{sec:interpretation}

This paper develops a control-theoretic interpretation of agency in
feedback systems by defining agency as runtime decision authority
	over elements of the control architecture. Under this interpretation,
the central question is not how control inputs are computed, but which
components of the closed-loop system may be modified during operation,
including controller parameters, controller families, workflow
composition, and control objectives.

A key implication is that increasing runtime authority alters the
class of dynamical system governing the closed loop. Parameter
adaptation produces time-varying dynamics, controller selection induces
switched systems, architectural reconfiguration leads to hybrid systems,
and endogenous design evolution introduces supervisory dynamics with
additional internal states. Thus, increasing agency corresponds to a
progression toward richer and more complex dynamical system classes.

The distinguishing feature of the proposed framework is that these
decision variables are treated as endogenous, generated by an
internal decision process, rather than as exogenous or
designer-specified signals as in classical robust, adaptive, or switched
control. As a result, stability must be analyzed jointly with the
dynamics that generate these decisions.

While classical control theory provides mature tools for analyzing
time-varying, switched, and hybrid systems, these tools are typically
applied under the assumption that the corresponding signals are
externally specified or satisfy prescribed constraints. In contrast,
agentic systems generate these signals internally, leading to
coupled dynamical mechanisms. The analysis in
Section~\ref{sec:stability} shows that such coupling introduces stability
conditions that do not arise when each mechanism is considered in
isolation.

The fully coupled result of Theorem~2 provides a conservative sufficient condition under which
adaptation, delay, switching, architectural reconfiguration, and
endogenous design evolution can be certified simultaneously. This result
formalizes the notion that all decision mechanisms consume a shared
stability margin.

\subsection{Design Implications for Agentic Control Systems}
\label{sec:design_implications}

The analysis of Section~\ref{sec:stability} leads to a unified set of
design principles for agentic control systems, in which stability
constraints are interpreted as constraints on the agent's decision
	process.

When the agent adapts controller parameters (Level~2), the adaptation
rate must satisfy $\|\Phi_\theta\| \le \rho$ with
$
\rho < \frac{\gamma}{L_\theta},
$
ensuring a positive effective decay margin
$\gamma_{\mathrm{eff}} = \gamma - L_\theta\rho > 0$ as required by
Theorem~1. In practice, this implies that
parameter updates should be filtered or rate-limited relative to the
plant's nominal stability margin.

When the agent selects among controller families (Level~3), switching
policies must satisfy dwell-time or hysteresis constraints. In the
presence of concurrent adaptation, the admissible dwell-time is given by
\begin{equation}
	\tau_a > \frac{\ln\nu}{\gamma - L_\theta\rho},
	\label{eq:design_dwell}
\end{equation}
which tightens as the adaptation rate increases. This highlights a
fundamental trade-off: rapid adaptation is incompatible with aggressive
switching. Decision rules that do not enforce
\eqref{eq:design_dwell}, including policies generated without explicit
stability awareness, may destabilize the closed loop.

When delays due to sensing, reasoning, or tool invocation are present,
they accumulate into the total delay bound $\bar{\tau}$, reducing the
effective decay rate. The corresponding dwell-time constraint becomes
\begin{equation}
	\tau_a > \frac{\ln\nu}{\gamma - \beta\bar{\tau}},
	\label{eq:design_delay}
\end{equation}
as shown in Proposition~1. This implies that
latency is not merely a performance consideration but a stability-critical
quantity: increased delay must be compensated by more conservative
switching behavior.

When architectural reconfiguration is permitted (Level~4), transitions
between configurations must satisfy bounded Lyapunov growth conditions
together with appropriate dwell-time constraints. More generally,
reconfiguration events should be treated as structured jumps whose
frequency and impact are explicitly regulated.

When endogenous design evolution is allowed (Level~5), the rate of design
updates must be bounded, as captured by the term $L_d\eta$ in
Theorem~2. In addition, admissibility constraints encoded in $\mathcal{G}_c$ must be enforced to ensure that generated
designs remain compatible with stability and safety requirements. For example, in safe reinforcement learning,
controller updates are often constrained using Lyapunov or control barrier function (CBF) conditions, ensuring
that each policy update preserves a certified decrease condition or forward invariance of a safe set.

Taken together, these conditions define a stability budget:
\begin{align}
	\label{eq:StabilityMargin}
\gamma >
L_\theta\rho + L_d\eta + \beta\bar{\tau}
+
\frac{\ln\nu_\sigma}{\tau_{a,\sigma}}
+
\frac{\ln\nu_c}{\tau_{a,c}},
\end{align}
which must be satisfied to guarantee stability. Continuous
mechanisms (adaptation, design evolution, delay) and discrete mechanisms
(switching, reconfiguration) therefore compete for a shared margin. This
interpretation provides a systematic way to allocate allowable rates,
latencies, and switching frequencies across different layers of the
agent.

\subsection{Design Mechanism: State-Dependent Hysteresis}
\label{sec:hysteresis}

The dwell-time requirement in Theorem~1 can be
enforced constructively through a hysteresis mechanism on the switching
logic. The following result shows that a state-dependent hysteresis band
induces a uniform lower bound on inter-switching times and its proof is provided in Appendix \ref{Appendix:3}.

\textbf{Proposition 2}:
	Consider the Level~2-3 closed-loop system of
	Theorem~1, and let the augmented continuous state be
	\[
	\varsigma := (x,\theta,m,z,r),
	\]
	evolving on a compact forward-invariant set \(\mathcal{X}\). Let
	\(\psi:\mathcal{X}\to\mathbb{R}\) be continuously differentiable, and define
	the scalar decision score
	\[
	\eta(t) := \psi(\varsigma(t)).
	\]
	Assume for simplicity that the switching signal \(\sigma(t)\in\{1,2\}\) is generated by the
	state-dependent hysteresis rule
	\[
	\sigma(t^+) =
	\begin{cases}
		1, & \eta(t) \ge h(\varsigma(t)),\\[2pt]
		2, & \eta(t) \le -h(\varsigma(t)),\\[2pt]
		\sigma(t^-), & |\eta(t)| < h(\varsigma(t)),
	\end{cases}
	\]
	where \(h:\mathcal{X}\to\mathbb{R}_{>0}\) is continuous. Let \(F\) denote the
	closed-loop vector field of the augmented continuous dynamics, and define
	\[
	\underline{h} := \inf_{\varsigma\in\mathcal{X}} h(\varsigma) > 0,
	\qquad
	\overline{M} := \sup_{\varsigma\in\mathcal{X}}
	\|\nabla_{\varsigma}\psi(\varsigma)\|\,\|F(\varsigma)\| < \infty.
	\]
	Then the following statements hold.
	\begin{enumerate}
		\item Every pair of consecutive switching instants \(t_k<t_{k+1}\) satisfies
		\[
		t_{k+1}-t_k \ge \frac{2\underline{h}}{\overline{M}} =: \tau_h.
		\]
		\item The switching signal admits uniform dwell-time \(\tau_h\), and hence
		satisfies an average dwell-time condition with \(\tau_a \ge \tau_h\).
		\item If
		\[
		\frac{2\underline{h}}{\overline{M}}
		>
		\frac{\ln\nu}{\gamma-L_\theta\rho},
		\]
		then the dwell-time condition of Theorem~1 is
		satisfied, and the origin is exponentially stable.
	\end{enumerate}

\begin{remark}
	A sufficient design guaranteeing the condition in
	Proposition~2(iii) is to choose
	\(h:\mathcal{X}\to\mathbb{R}_{>0}\) such that
	\[
	h(\varsigma)
	>
	\frac{\overline{M}}{2}\cdot
	\frac{\ln\nu}{\gamma-L_\theta\rho}
	\qquad \forall\,\varsigma\in\mathcal{X}.
	\]
	Then \(\underline{h}\) satisfies the required bound, and the hysteresis rule
	enforces a dwell-time compatible with Theorem~1. More
	generally, allowing \(h(\varsigma)\) to vary over \(\mathcal{X}\) can reduce
	unnecessary switching in regions where the decision score evolves more rapidly
	or more slowly, while stability is still certified through the global lower
	bound \(\underline{h}\).
\end{remark}

\subsection{Limitations and Future Directions}
\label{sec:limitations}

Several limitations of the present work should be noted. While the
analysis is formulated in a general nonlinear setting, verification of
the assumptions (e.g., bounds on $L_\theta$, $L_d$, and delay effects)
may be nontrivial for complex decision processes such as those involving
large language models. Providing systematic methods to estimate or bound
these quantities remains an important direction for future work.

The fully coupled result of Theorem~2 provides a
conservative sufficient condition for stability. While it establishes
that all mechanisms can be analyzed within a unified Lyapunov framework,
it does not provide a tight or necessary characterization of the
stability region. Reducing conservatism and deriving less restrictive
conditions is an open problem.

The present formulation considers a single agent embedded in a feedback
loop. Extending the framework to multi-agent systems introduces
additional challenges related to interaction, coordination, and
information asymmetry. As noted in Remark~\ref{rem:mas}, the stability
conditions may depend on quantities (e.g., adaptation rates, switching
policies) that are controlled by different agents, making them difficult
to certify in a decentralized setting. Developing distributed stability
conditions for such systems is a key direction for future research.

Finally, the role of semantic uncertainty in the information signals
driving decision updates is not explicitly modeled. In particular,
interpretation errors in language-based or perception-based inputs may
affect switching, adaptation, or goal generation in ways that are not
captured by the current framework. Incorporating such effects into the
stability analysis represents another important extension of this work.

\section{Simulations}
\label{sec:simulations}

This section provides numerical validation of the stability
conditions developed in Section~\ref{sec:stability}. The
simulations are designed to (i)~isolate the effects of
individual mechanisms, switching, delay, and structural
reconfiguration, and (ii)~verify the coupled stability
budget of Theorem~2. All examples use low-order systems so
that the observed behavior can be directly interpreted in
relation to the theoretical bounds. Matrix values and
controller parameters are provided in Appendix~A.

\subsection{Simulation Setup}

We consider the second-order switched linear time-delay system
\begin{equation}
	\label{eq:Sim_section_augmented}
	\dot{x}(t)=A_{\sigma(t)}x(t)+A_d\,x(t-\bar{\tau}),
	\qquad \sigma(t)\in\{1,2\},
\end{equation}
with $
	A_1=
	\begin{bmatrix}
		-3.8897 & -3.2679\\
		1.5381 & 0.7197
	\end{bmatrix},
	A_2=
	\begin{bmatrix}
		-3.1172 & 0.4840\\
		-5.4957 & 0.4516
	\end{bmatrix},
	A_d=
	\begin{bmatrix}
		0 & 0\\
		0.8 & 0.2
	\end{bmatrix}$.
Both modes \(A_1\) and \(A_2\) are individually stable. The switching
signal is generated endogenously from the scalar decision score
\(\eta(t)=x_1(t)\), using either a threshold rule or a hysteresis-based
policy, so that the closed loop captures Level~3 strategy switching with
decision delay and dwell-time effects.

To study the interaction between switching, delay, and adaptation, the
controller parameter \(\theta(t)\) evolves according to
\begin{equation}
	\dot{\theta}(t)=\rho\big(\theta^\ast(t)-\theta(t)\big),\qquad
	\theta^\ast(t)=0.3\tanh(\|x(t)\|),
\end{equation}
where \(\rho>0\) is the adaptation rate. This provides a simple Level~2
mechanism whose coupling with switching and delay is used to evaluate the
stability budget in Theorem~2.

For the structural reconfiguration study (Level~4), we additionally
consider a second-order plant implemented under two alternative control
architectures: (i) direct state feedback and (ii) an observer-based
realization. Switching between these architectures changes the internal
signal flow and introduces mode-dependent internal dynamics, thereby
producing a hybrid closed-loop system with reconfiguration events.

\subsection{Mechanism-Specific Stability Effects}

\subsubsection{Switching, Dwell-Time, and Delay
	(Theorem~1 and Proposition~1)}

\begin{figure}[t]
	\centering
	\includegraphics[width=\linewidth]{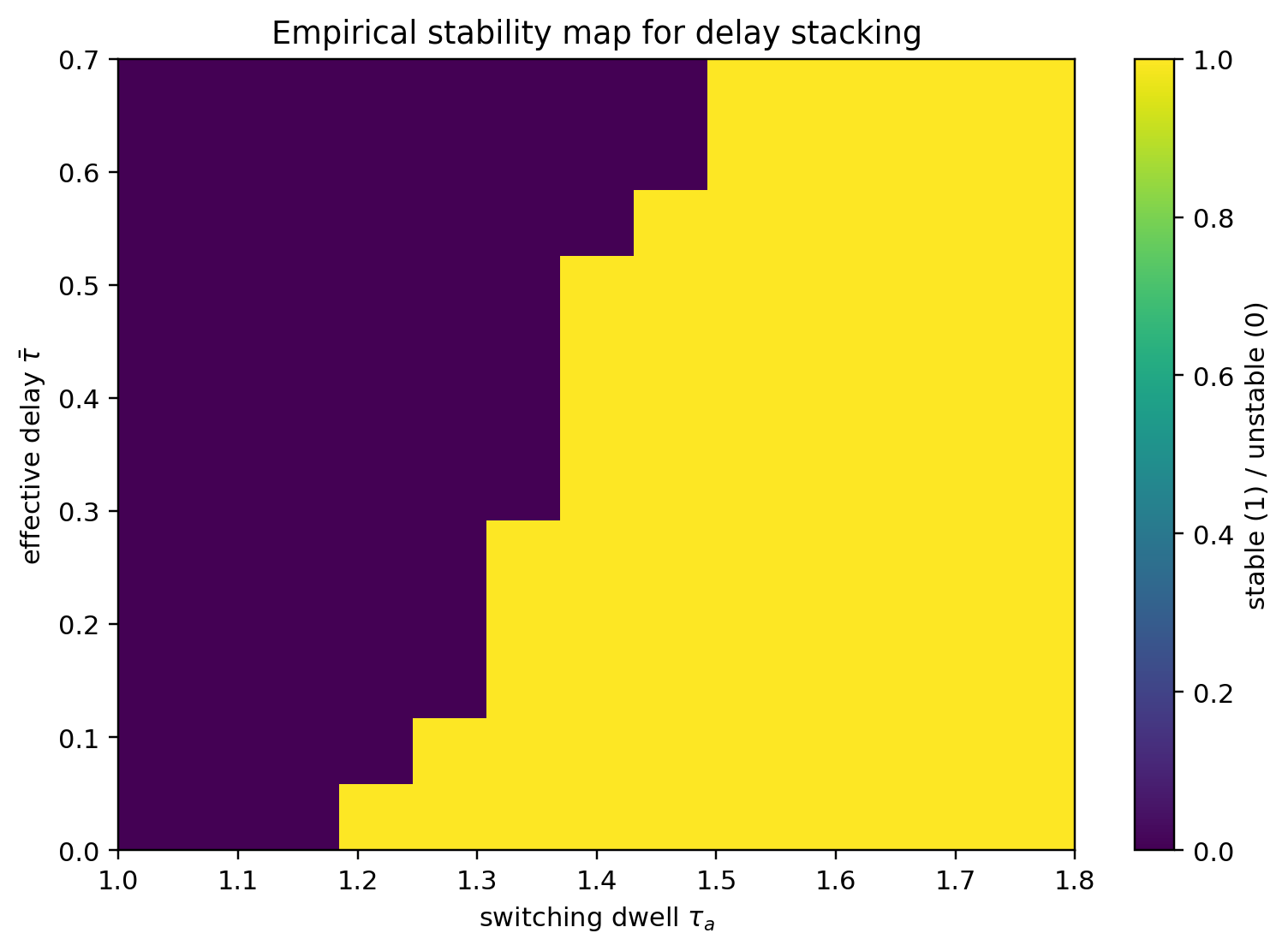}
	\caption{Empirical stability map over dwell-time $\tau_a$
		and delay $\bar{\tau}$. At $\bar{\tau}=0$, the boundary
		falls near $\tau_a\approx 1.35\,\mathrm{s}$, close to
		the theoretical threshold $\tau_a^*=1.30\,\mathrm{s}$
		of Theorem~1. As delay increases, the boundary shifts
		monotonically, consistent with the effective decay rate
		$\gamma_{\mathrm{eff}}=\gamma-\beta\bar{\tau}$ of
		Proposition~1.}
	\label{fig:delay_map}
\end{figure}

Theorem~1 predicts stability when the average dwell-time
satisfies
\begin{equation}
\tau_a > \tau_a^* := \frac{\ln\nu}{\gamma - L_\theta\rho}.
\end{equation}
In the absence of adaptation ($\rho=0$), the system
parameters give $\tau_a^* = \ln\nu/\gamma = 1.30\,\mathrm{s}$.
The empirical stability boundary, visible in
Fig.~\ref{fig:delay_map} at $\bar{\tau}=0$, falls near
$\tau_a\approx 1.2\,\mathrm{s}$, consistent with this
prediction.

Proposition~1 further predicts that delay reduces the
effective decay rate, shifting the stability boundary to
\begin{equation}
\tau_a > \frac{\ln\nu}{\gamma - \beta\bar{\tau}}.
\end{equation}
Fig.~\ref{fig:delay_map} shows the full empirical stability
map over $(\tau_a,\bar{\tau})$. The boundary shifts
monotonically with delay as the delay increases, confirming
that delay and switching jointly consume the available
stability margin rather than acting independently.

\subsubsection{Structural Reconfiguration (Level~4)}

\begin{figure}[t]
	\centering
	\includegraphics[width=\linewidth]{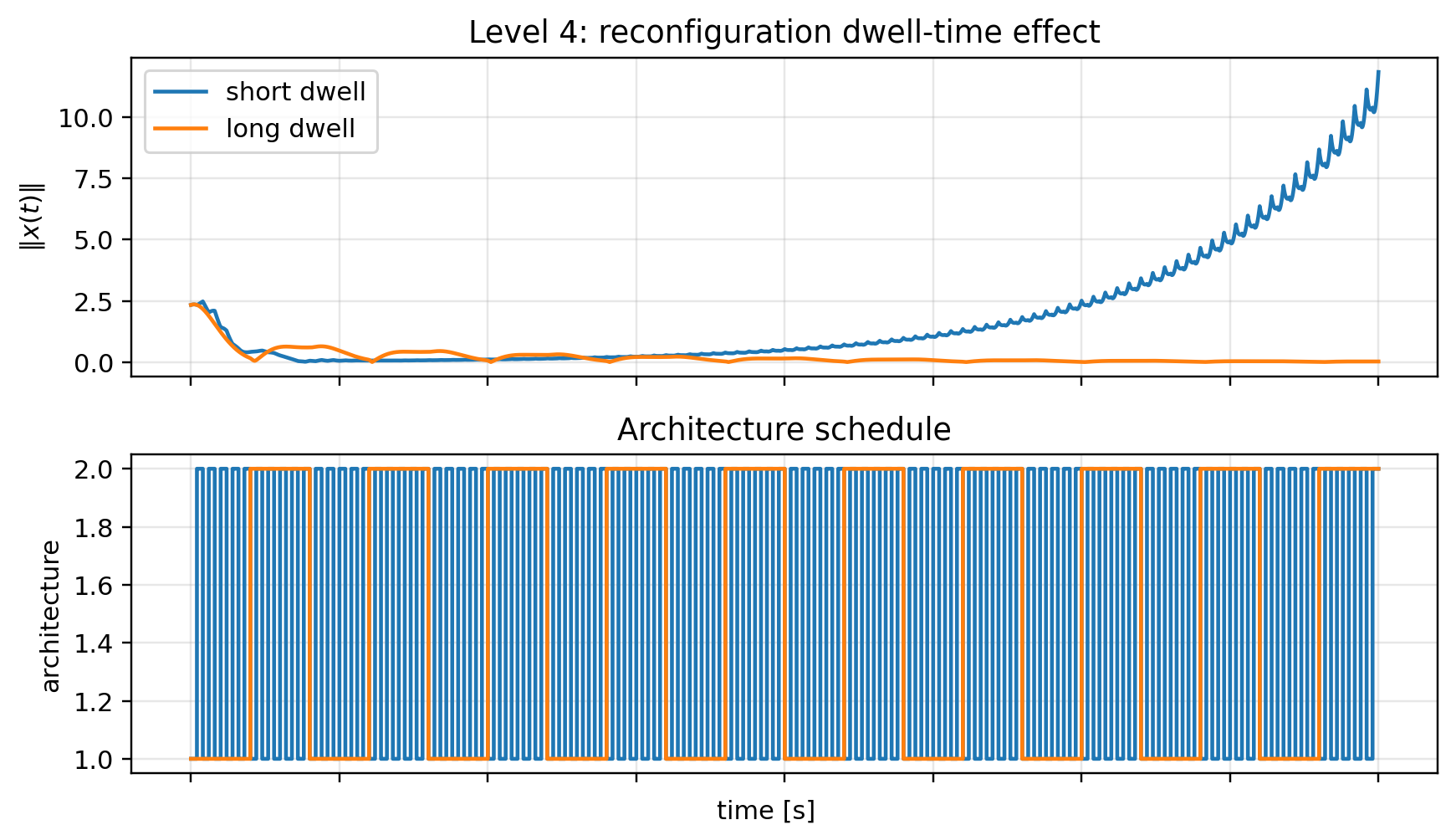}
	\caption{Effect of structural reconfiguration. Frequent
		switching destabilizes the system through accumulated
		jump effects, consistent with the jump comparability
		condition of Theorem~2.}
	\label{fig:level4}
\end{figure}

Fig.~\ref{fig:level4} shows the effect of switching between
control architectures. Although each architecture is
individually stabilizing, frequent reconfiguration induces
instability through repeated excitation of internal states,
while infrequent switching preserves stability. This is
consistent with the jump comparability condition of
Theorem~2,
\begin{equation}
V^+(x) \leq \nu\,V^-(x),
\end{equation}
which introduces a multiplicative growth factor at each
reconfiguration instant that must be offset by sufficient
dwell time.

\subsection{Fully Coupled Stability Budget (Theorem~2)}

We now consider the combined effect of adaptation, switching,
and delay operating simultaneously. The system evolves
according to~\eqref{eq:Sim_section_augmented} with adaptive
parameter dynamics
\begin{equation}
\dot{\theta}(t)
= \rho\,\bigl(\theta^*(t)-\theta(t)\bigr),
\qquad
\theta^*(t) = 0.3\tanh\!\bigl(\|x(t)\|\bigr),
\label{eq:theta_dyn2}
\end{equation}
so that $\theta(t)$ tracks a state-dependent target at rate
$\rho$. The switching signal is generated from
$\mathrm{sign}(x_1(t))$, subject to a dwell-time constraint
$\tau_a$. The Lyapunov constants are estimated from the
spectral properties of $A_1$ and $A_2$: $\gamma=0.609$
(least stable eigenvalue of $A_2$), $\nu=2.2$,
$L_\theta=0.8$, and $\beta=2.5$.

The stability condition of Theorem~2 reduces to the budget
\begin{equation}
\lambda = \gamma - L_\theta\rho
- \beta\bar{\tau}
- \frac{\ln\nu}{\tau_a},
\label{eq:budget_sim}
\end{equation}
where $\rho$ is the constant adaptation rate
in~\eqref{eq:theta_dyn2}. Theorem~2 guarantees exponential
stability when $\lambda>0$. In the stable case
($\rho=0.15$, $\tau_a=4.0\,\mathrm{s}$,
$\bar{\tau}=0.03\,\mathrm{s}$), the budget evaluates to
$\lambda=0.609-0.120-0.075-0.197=+0.217>0$. In the
unstable case ($\rho=3.5$,
$\tau_a=0.4\,\mathrm{s}$, $\bar{\tau}=0.20\,\mathrm{s}$),
it evaluates to $\lambda=0.609-2.800-0.500-1.971=-4.662<0$.

\begin{figure}[t]
	\centering
	\includegraphics[width=\linewidth]{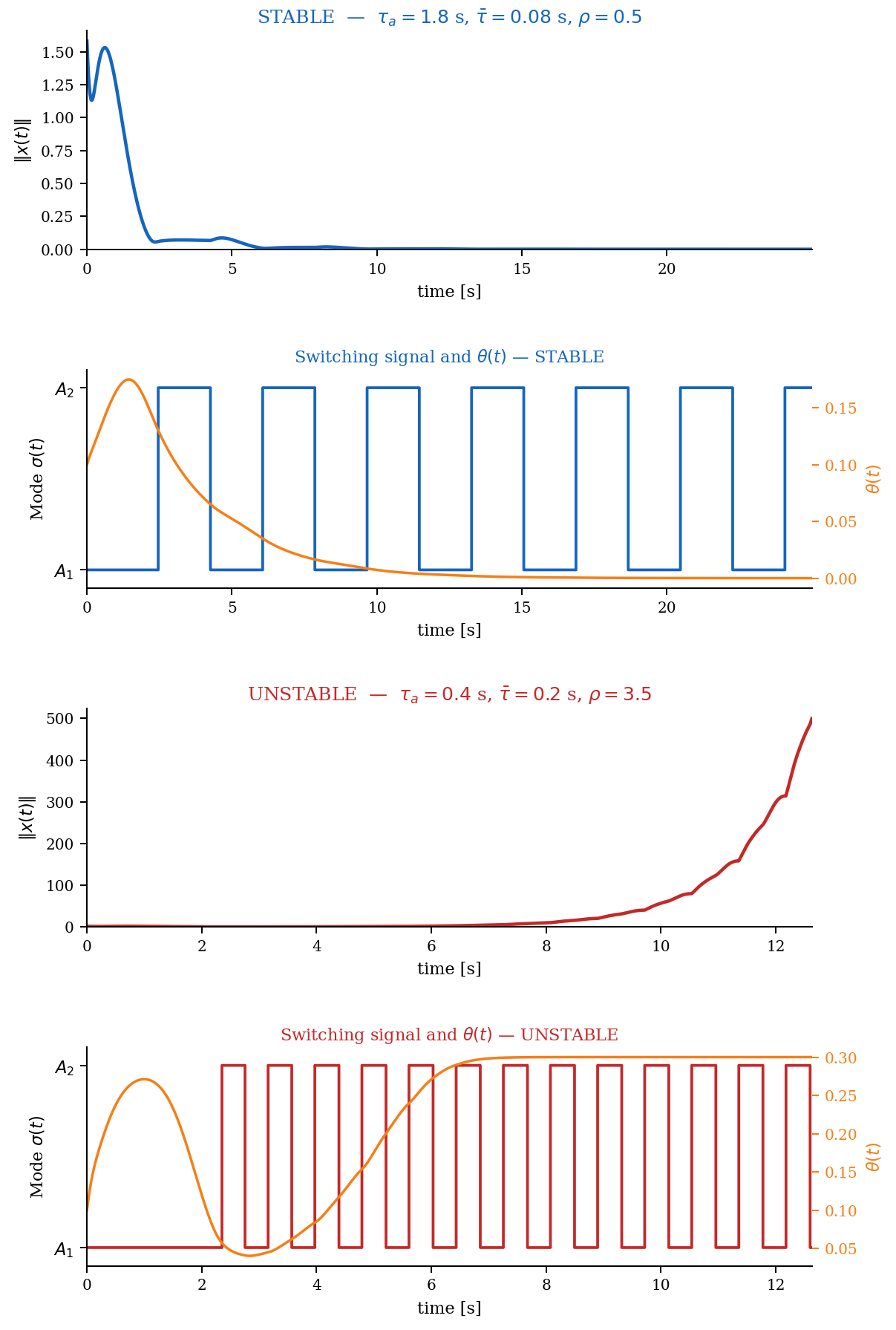}
	\caption{State norm $\|x(t)\|$, switching signal
		$\sigma(t)\in\{A_1,A_2\}$, and adaptive parameter
		$\theta(t)$ for the stable (top) and unstable (bottom)
		cases. Each switch between $A_1$ (regulation) and
		$A_2$ (tracking) constitutes a simultaneous Level~4
		reconfiguration event. This figure illustrates the importance of enforcing dwelling time constraints.}
	\label{fig:sim_state_mode}
\end{figure}

Fig.~\ref{fig:sim_state_mode} shows the state norm
$\|x(t)\|$, switching signal $\sigma(t)$, and adaptive
parameter $\theta(t)$ for both cases. The stable case
converges to the origin; the unstable case diverges. Here,
$A_1$ and $A_2$ represent the regulation and tracking
closed-loop architectures, respectively, so each switch is
simultaneously a controller selection and a structural
reconfiguration. The single jump-ratio term $\ln\nu/\tau_a$
in~\eqref{eq:budget_sim} therefore captures both
contributions, directly instantiating the coupled condition
of Theorem~2.

\begin{figure}[t]
	\centering
	\includegraphics[width=\linewidth]{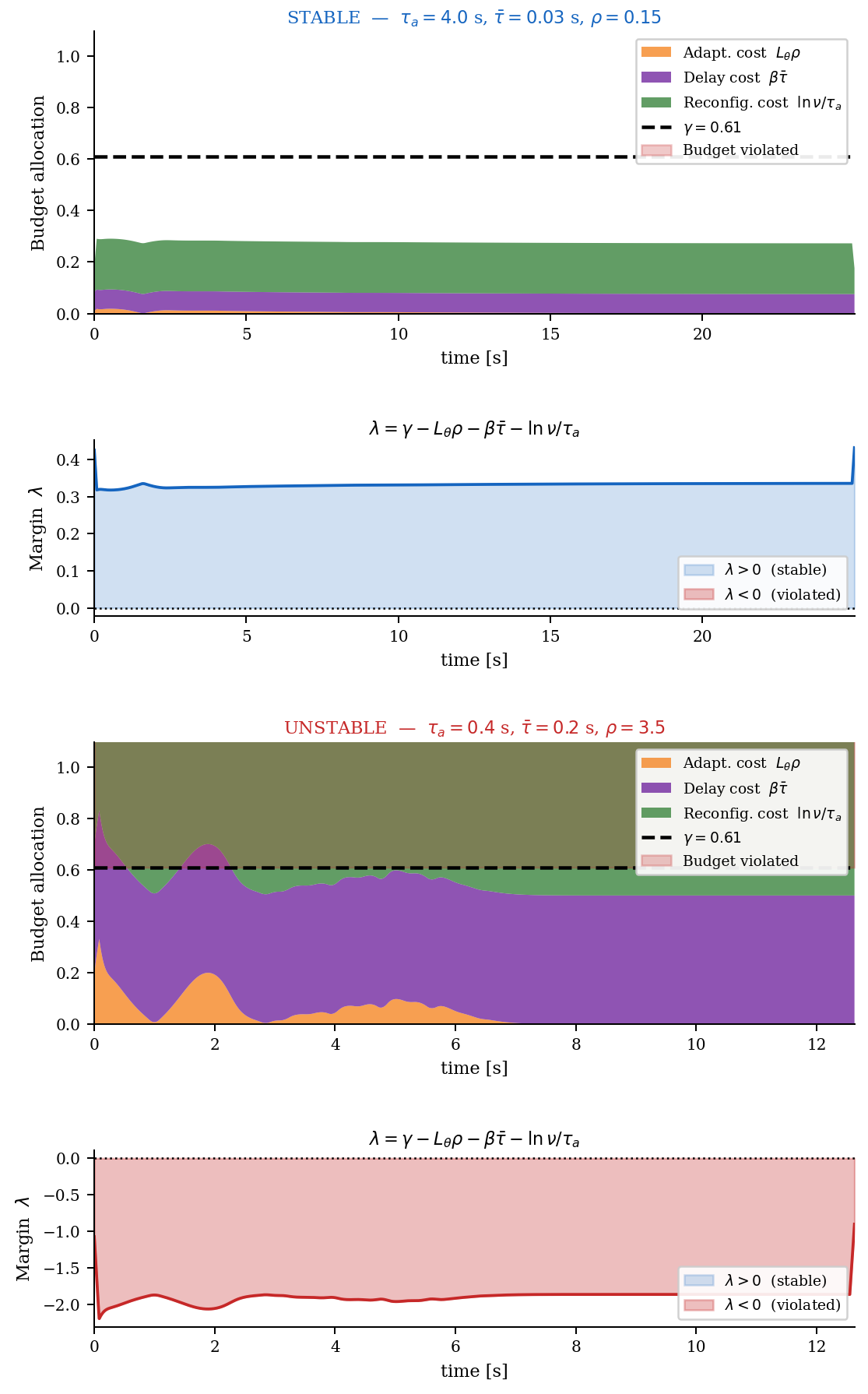}
	\caption{Real-time stability budget~\eqref{eq:budget_sim}
		for the stable (top) and unstable (bottom) cases.
		\textit{Upper two panel}: stacked cost terms
		$L_\theta\rho$ (adaptation), $\beta\bar{\tau}$ (delay),
		and $\ln\nu/\tau_a$ (reconfiguration) against the total
		budget $\gamma$ (dashed line); shading marks where the
		budget is exceeded. \textit{Lower two panel}:
		effective decay margin $\lambda$. In the stable case
		$\lambda>0$ throughout; in the unstable case $\lambda<0$
		from the outset, predicting divergence ahead of any
		visible growth in $\|x(t)\|$.}
	\label{fig:sim_budget}
\end{figure}

Fig.~\ref{fig:sim_budget} shows the corresponding real-time
stability budget. In the stable case the three cost terms
remain below $\gamma$ and $\lambda>0$ throughout. In the
unstable case the stacked costs exceed $\gamma$ from the
outset, with $\lambda<0$ predicting divergence before any
visible growth in $\|x(t)\|$ is apparent. This confirms
that the sign of $\lambda$ serves as a reliable leading
indicator of instability in the fully coupled setting.

\section{Conclusion and Future Work}
This paper presented a control-theoretic framework for modeling agentic AI systems embedded within feedback control loops. In the proposed formulation, agency is interpreted as decision authority over elements of the control architecture, including controller parameters, goal descriptors, tool invocation, and controller structure. By embedding capabilities such as memory, learning, interaction, and tool use into a unified dynamical representation, the framework provides a systematic way to interpret agentic AI behavior using familiar concepts from control theory. The analysis shows that increasing agency changes the induced closed-loop dynamics: adaptation yields time-varying systems, strategy selection leads to switching, reasoning and tool use introduce delay, and structural reconfiguration produces hybrid dynamics, with generative agency enabling evolving objectives. While these mechanisms are classical in isolation, the proposed framework shows that they arise jointly when the decision process is endogenous to the feedback loop, bridging agentic AI capabilities with established dynamical systems analysis.

Several directions remain for future research. First, the present work focuses primarily on modeling and illustrative analysis. A natural next step is the development of more complete stability and performance guarantees for agentic control architectures, including conditions under which parameter adaptation, controller switching, architectural reconfiguration, or objective updates preserve closed-loop stability. Second, the framework should be extended to multi-agent settings in which multiple agentic controllers interact with one another and with shared environments. Such systems may exhibit coordination dynamics, communication delays, and emergent behaviors that require new analysis tools from distributed control and multi-agent systems. One can also extend the idea of stability budget to the case of multi-agent systems and use the notions of consensus on the stability budget as a way to ensure stability over the multi-agent system.
Third, practical design guidelines are needed for incorporating agentic AI into safety-critical control systems. Examples include limiting adaptation rates, imposing dwell-time constraints on controller switching, bounding decision-induced delays, and enforcing governance constraints on objective or architecture synthesis. Additional directions include studying robustness of agentic controllers under model uncertainty, analyzing the effects of semantic ambiguity in interaction signals, and developing verification and certification methods for AI-driven control architectures.

\appendices
\section{Proof of Theorem 1}
\label{Appendix:1}
\begin{proof}
	Along a flow interval in mode \(p\), the derivative of \(V_p\) satisfies
	\begin{align}
	\dot{V}_p
	&=
	\frac{\partial V_p}{\partial x}(x,\theta)\,f_p(x,\theta)
	+
	\frac{\partial V_p}{\partial\theta}(x,\theta)\cdot\dot{\theta}
	\nonumber\\
	&\le
	-\gamma V_p(x,\theta)
	+
	\left|
	\frac{\partial V_p}{\partial\theta}(x,\theta)\,
	\Phi_\theta(\theta,I)
	\right|
	\nonumber\\
	&\le
	-\gamma V_p + L_\theta\rho\,V_p
	=
	-(\gamma-L_\theta\rho)\,V_p.
	\end{align}
	Thus \(V_p\) decays exponentially along each flow interval with effective
	rate
	\[
	\gamma_{\mathrm{eff}}:=\gamma-L_\theta\rho>0.
	\]
	At each switching instant, Assumption~3 gives
	\[
	V_{\sigma^+}(x,\theta)\le\nu\,V_{\sigma^-}(x,\theta).
	\]
	Combining flow decay and jump growth over \([t,t+T]\) yields
	\[
	V\bigl(x(t+T),\theta(t+T)\bigr)
	\le
	\nu^{N_\sigma(t,t+T)}
	e^{-\gamma_{\mathrm{eff}}T}
	V\bigl(x(t),\theta(t)\bigr).
	\]
	Applying the average dwell-time bound gives
	\[
	V\bigl(x(t+T),\theta(t+T)\bigr)
	\le
	\nu^{N_0}
	\exp\!\left(
	-\Bigl(\gamma_{\mathrm{eff}}-\tfrac{\ln\nu}{\tau_a}\Bigr)T
	\right)
	V\bigl(x(t),\theta(t)\bigr).
	\]
	Under~\eqref{eq:cross_condition}, the exponent is strictly negative, which
	implies exponential convergence of \(x(t)\). Boundedness of \(\theta(t)\)
	follows from compact forward invariance of \(\Theta\). \hfill
\end{proof}
\section{Proof of Theorem 2}
\label{Appendix:2}
\begin{proof}
	Let \(V(t):=V_{p(t)}(\chi(t))\). Since the hybrid system is Zeno-free, the
	solution can be partitioned into flow intervals \([t_k,t_{k+1})\) separated
	by jump instants.
	
	Along any flow interval with constant mode \(p\),
	\[
	\dot{V}_p
	=
	\frac{\partial V_p}{\partial \bar{\xi}}F_p(\bar{\xi};\theta,d)
	+
	\frac{\partial V_p}{\partial \theta}\Omega_\theta(\theta,I)
	+
	\frac{\partial V_p}{\partial d}\Omega_d(d,I)
	+
	\Delta_\tau V_p.
	\]
	Using \textnormal{(A2)}-\textnormal{(A5)} gives
	\[
	\dot{V}_p \le -\lambda V_p.
	\]
	Hence, for any flow subinterval \([s,r)\),
	\[
	V(r^-)\le e^{-\lambda(r-s)} V(s^+).
	\]
	
	At each jump time \(t_k\), condition \textnormal{(A6)} gives
	\[
	V(t_k^+)
	\le
	\nu_\sigma^{\delta_{\sigma,k}}
	\nu_c^{\delta_{c,k}}
	V(t_k^-).
	\]
	
	Combining flow decay and jump growth over \([0,T]\) yields
	\[
	V(T)
	\le
	\nu_\sigma^{N_\sigma(0,T)}
	\nu_c^{N_c(0,T)}
	e^{-\lambda T} V(0).
	\]
	Applying the average dwell-time bounds \textnormal{(A7)},
	\[
	V(T)
	\le
	\nu_\sigma^{N_{0,\sigma}}
	\nu_c^{N_{0,c}}
	\exp\!\left[
	-\left(
	\lambda
	-
	\frac{\ln \nu_\sigma}{\tau_{a,\sigma}}
	-
	\frac{\ln \nu_c}{\tau_{a,c}}
	\right)T
	\right] V(0).
	\]
	
	Define
	\[
	\mu :=
	\lambda
	-
	\frac{\ln \nu_\sigma}{\tau_{a,\sigma}}
	-
	\frac{\ln \nu_c}{\tau_{a,c}} > 0.
	\]
	Then
	\[
	V(T)\le C e^{-\mu T} V(0),
	\qquad
	C := \nu_\sigma^{N_{0,\sigma}} \nu_c^{N_{0,c}}.
	\]
	
	Using the bounds in \textnormal{(A1)},
	\[
	\underline{\alpha}\|\chi(T)\|^2
	\le V(T)
	\le C \overline{\alpha} e^{-\mu T} \|\chi(0)\|^2.
	\]
	Thus,
	\[
	\|\chi(T)\|
	\le
	\sqrt{\frac{C\overline{\alpha}}{\underline{\alpha}}}
	\, e^{-\mu T/2} \|\chi(0)\|.
	\]
	
	The result follows with
	\[
	K := \sqrt{\frac{C\overline{\alpha}}{\underline{\alpha}}},
	\qquad
	\omega := \frac{\mu}{2}.
	\]
\end{proof}
\section{Proof of Proposition 2}
\label{Appendix:3}
\begin{proof}
	Since \(\psi\) is continuously differentiable and \(\varsigma(\cdot)\) evolves
	under the closed-loop vector field \(F\), the chain rule gives
	\[
	\dot{\eta}(t)
	=
	\nabla_{\varsigma}\psi(\varsigma(t))\cdot F(\varsigma(t))
	\]
	for almost all \(t\). By continuity of \(\nabla_{\varsigma}\psi\) and \(F\) on
	the compact set \(\mathcal{X}\),
	\[
	|\dot{\eta}(t)|
	\le
	\|\nabla_{\varsigma}\psi(\varsigma(t))\|\,\|F(\varsigma(t))\|
	\le
	\overline{M}
	\]
	for almost all \(t\).
	
	Let \(t_k<t_{k+1}\) be consecutive switching instants. Without loss of
	generality, suppose the switch at \(t_k\) places the system in mode \(1\). By
	the hysteresis rule,
	\[
	\eta(t_k)\ge h(\varsigma(t_k))\ge \underline{h}.
	\]
	The next switch cannot occur until the decision score reaches the opposite
	threshold, that is,
	\[
	\eta(t_{k+1})\le -h(\varsigma(t_{k+1}))\le -\underline{h}.
	\]
	Hence,
	\[
	|\eta(t_{k+1})-\eta(t_k)| \ge 2\underline{h}.
	\]
	Since \(\eta(\cdot)\) is absolutely continuous,
	\[
	\eta(t_{k+1})-\eta(t_k)=\int_{t_k}^{t_{k+1}}\dot{\eta}(s)\,ds,
	\]
	so
	\[
	2\underline{h}
	\le
	\int_{t_k}^{t_{k+1}} |\dot{\eta}(s)|\,ds
	\le
	\overline{M}(t_{k+1}-t_k).
	\]
	Therefore,
	\[
	t_{k+1}-t_k \ge \frac{2\underline{h}}{\overline{M}} = \tau_h,
	\]
	which proves (i). Statement (ii) follows because uniform dwell-time implies
	average dwell-time with \(\tau_a\ge\tau_h\). Finally, if
	\[
	\tau_h=\frac{2\underline{h}}{\overline{M}}
	>
	\frac{\ln\nu}{\gamma-L_\theta\rho},
	\]
	then the dwell-time condition required by
	Theorem~1 holds, and exponential stability follows.
\end{proof}
\bibliographystyle{IEEEtran}
\bibliography{sample}  

\end{document}